# WHAT "CROWDSOURCING" OBSCURES: EXPOSING THE DYNAMICS OF CONNECTED CROWD WORK DURING DISASTER

Kate Starbird

University of Colorado Boulder
430 UCB
Boulder, CO 80309-0430
catharine.starbird@colorado.edu

**ABSTRACT**

The aim of this paper is to demonstrate that the current understanding of *crowdsourcing* may not be broad enough to capture the diversity of crowd work during disasters, or specific enough to highlight the unique dynamics of information organizing by the crowd in that context. In making this argument, this paper first unpacks the crowdsourcing term, examining its roots in open source development and outsourcing business models, and tying it to related concepts of human computation and collective intelligence. The paper then attempts to characterize several examples of crowd work during disasters using current definitions of crowdsourcing and existing models for human computation and collective intelligence, exposing a need for future research towards a framework for understanding crowd work.

**INTRODUCTION**

Sociologists of disaster have repeatedly shown that the first responders to disasters are rarely the formal organizations charged to respond, but are instead spontaneous volunteers who converge on the scene and begin to help (Fritz & Mathewson, 1957; Dynes, 1970; Palen & Liu, 2007). People now use ICTs, including social media, to converge digitally during and after disaster events (e.g. Hughes & Palen, 2009; Qu et al., 2009), and this convergence via social media has led to appearance of a new kind of volunteer: the digital volunteer.

During the early aftermath of the 2010 Haiti earthquake, groups of digital volunteers mobilized to provide assistance to those affected and those responding. The Disaster Relief 2.0 report (Harvard Humanitarian Initiative, 2011) described several of those efforts, including how "crisis mapping" volunteers came together to create and improve publicly available maps of the affected areas[1], as well as how a group of volunteers organized around an Ushahidi instance to collect, process, verify, and map citizen reports of damage and immediate needs[2]. Additionally, hundreds of Twitter users "self-deployed" on that platform and attempted to assist in a number of capacities: by rebroadcasting what they thought was "good information," verifying information, routing information, and matching needs with offers of help (Starbird & Palen, 2011).

Throughout the early response and relief period, popular media and the digerati praised the role of "crowdsourcing" (e.g. Biewald & Leila, 2010; Mullins, 2010). The Disaster Relief 2.0 Report also used that term—28 times—to characterize the activities of digital volunteer groups during that event. Crowdsourcing efforts have been credited as serving important roles in response efforts for numerous crisis events since. However, the widespread use of this popular term, employed as a blanket descriptor for a variety of activities, may be obscuring the complexity of human behaviors and computational systems that support them.

**UNPACKING CROWDSOURCING**

Crowdsourcing, in its broadest sense, involves leveraging the capabilities of a connected crowd to complete work. The term was coined by Howe (2006) who used it to describe a wide range of behaviors, tasks, and work practices that were newly enabled by Internet connectivity, including crowd brainstorming, crowd voting, collaborative creation, the division of complex tasks into micro-tasks for geographically dispersed workers, citizen science, collaborative filters and more (Howe, 2008). Closely related to both collective intelligence (Hiltz & Turoff, 1978) and human computation (von Ahn, 2005), crowdsourcing can be conceptualized as an extension of open-source projects and organizing techniques to non-programmer tasks and projects. Howe provided an umbrella term to a technological turn—a turn toward collective activity mediated by the Internet and its offspring (social media). The term itself has recently become a popular one, masking much of the diversity (of activity, task, work structure, work practices, etc.) that lies beneath it.

---

[1] OpenStreetMap Haiti: http://haiti.openstreetmap.nl/

[2] Ushahidi map: http://haiti.ushahidi.com/

### Outsourcing

The roots of crowdsourcing lie in both *outsourcing* and *open source*. Outsourcing is a strategy used by companies to utilize workers outside their employee base. Harland et al. (2005), define outsourcing as a company going outside its internal workforce to procure something, a service or product, that the company either previously performed or created in-house or could have performed or created in-house.

Crowdsourcing imitates outsourcing by utilizing external, often remote, workforces. However, outsourced workforces are typically found remote to the company utilizing them, but within another organization structure and often have co-located employees. Workers in crowdsourced projects can be dispersed across the entire globe, relying on their own resources (an Internet connection) to complete their work. Their only connections to other workers—if connected at all—are in many cases from within the crowdsourced project. Also, because they can involve thousands of participants in non-traditional work situations (in some cases performing a few minutes or even seconds of work), crowdsourcing projects are not always endeavors that could be performed in-house by an existing company, as outsourcing arrangements are.

### Open Source

Open source projects are software development efforts organized by programmers where, in many cases, the end project belongs to the public domain, not to established companies or even to the programmers themselves. The Linux project, established by Linus Torvald in 1992, is often cited as an ideal example of open source. That effort, as most open source projects since, utilized the Internet to recruit programmers and organize their work.

The key difference between open source and crowdsourcing is the skill set of the envisioned labor force: open source projects primarily incorporate computer programmers, while crowdsourcing projects can take advantage of non-programmers capable of a wide range of tasks requiring human intelligence. Open sourcing can be seen as the original form of crowdsourcing, which has now recognized potential application in other domains, though crowdsourcing has analogs that preexist open source, including application in the accessibility domain (Bigham, Ladner & Brodin, 2011).

### Collective Intelligence

Collective intelligence has been applied to the interpretation of a variety of phenomena in different domains. It has been offered as an explanation for seemingly intelligent behavior in large-scale interaction (Malone et al., 2009) for both humans and non-humans (Surweicki, 2005). It includes both conscious and instinctive—or spontaneous—collectively intelligent activity. Though collectively intelligent behavior in humans can be observed offline, the connected interaction enabled by Internet technologies has opened up new ways for the phenomenon to both occur and to be observed (Surowiecki, 2005). This paper considers collective intelligence enabled by computer mediated communication technology, on-line, focusing on the ability for connected and collaborating human beings to engage in collective problem-solving activities.

In its recent, popular manifestation, crowdsourcing leans on a definition of collective intelligence presented by Surowiecki as "the wisdom of crowds" (2005). Surowiecki outlines three types of problems that crowds were capable of solving: cognition, coordination, and cooperation. For problems of cognition, building off an understanding of the Delphi effect (see Hiltz & Turoff, 1978), Surowiecki argues that the wisdom of crowds manifests when there is a method of aggregating disparate views from diverse members of a crowd of sufficient size. These three components—aggregation, diversity, and large number of views—are key components of many of the crowdsourcing efforts outlined by Howe (2006, 2008). Coordination problems, which Surowiecki illustrated with examples of pedestrians navigating through crowded sidewalks, are, he claims, better addressed by the self-organizing, bottom-up solutions of crowd interaction than by top-down mandates.

### Human Computation

Crowdsourcing can also be approached from a perspective of human computation, another emerging field. Human computation has been defined as "a paradigm for utilizing human processing power to solve problems that computers cannot yet solve" (von Ahn, 2005, p. 3).

Human computation projects aim to capture large numbers of human judgments, using a variety of incentive strategies, to solve complex problems. Though early human computation projects focused on using humans to train computational algorithms, researchers eventually began to use the collective work of humans to solve problems directly. Human computation systems are now used for a wide range of applications, including von Ahn's reCAPTCHA project (2008), which uses word recognition tasks, required by some Internet sign-up portals to assert humanness, to do text translation (2008), and Amazon's popular Mechanical Turk platform which uses financial incentives to recruit crowd workers to other tasks provided by other users.

**EVOLVING FRAMEWORKS FOR CROWD WORK**

There are conflicting models for illustrating the relationships between concepts of crowdsourcing, human computation, and collective intelligence. Attempting to build a classification structure for the field of human computation, Quinn and Bederson (2011) position collective intelligence as an overarching category, encompassing all of crowdsourcing, most of human computation, and the portion of what they call "data mining" that is collaborative filtering. Attempting to differentiate between human computation and other forms of collective intelligence, they admit a small overlap between crowdsourcing and human computation, but assert that an important difference exists: "Whereas human computation replaces computers with humans, crowdsourcing replaces traditional human workers with members of the public" (p. 3). That demonstrates a more limited view of crowdsourcing than Howe's (2008), pushing the definition of crowdsourcing towards out-sourcing. It also places crowd-driven collaborative creation projects, like Wikipedia and open source, outside the realm of human computation. Quinn and Bederson also claim that crowdsourcing is separate from collaborative filtering, again conflicting with Howe's (2008) early definition, which included collaborative filtering—he uses examples of Amazon's recommendation system and the Google search engine—as part of crowdsourcing.

Malone et al. (2009) offer a framework for collective intelligence and characterize almost all of the examples of crowdsourcing offered by Howe (2006; 2008) within that umbrella term. In classifying systems further, they identify four building blocks (or "genes") of systems that leverage collective intelligence, asking: 1) *Who* does the work? Is it done within an organizational hierarchy or by the crowd? 2) *Why* are they doing it? What motivates them: money, love, or glory? 3) *What* are they accomplishing? Are they creating something or deciding something? 4) *How* is it being done, independently or dependently? However, many of the "crowdsourcing" efforts in the disaster domain exhibit characteristics along multiple dimensions within each block, e.g. motivations of love and glory with work done independently and dependently.

An examination of several models (Howe, 2008; Malone et al, 2009; Quinn & Bederson, 2011) reveals an evolving and incomplete understanding of how to characterize the crowd activities that are taking place on social media and the crowd leveraging systems that are being built to harness this work. These frameworks for collective intelligence and human computation both assume a "system" approach to collective intelligence within the connected crowd, excluding from their examples evidence of bottom-up, self-organizing efforts of crowd computation.

**CROWDSOURCING ≠ MICROWORK**

Another issue that risks obscuring the diversity of crowd work is the drift by researchers to equate crowdsourcing with Amazon's Mechanical Turk (mTurk) and other microwork systems. Mechanical Turk is a web-based platform that supports a microwork market, allowing project owners to distribute work across a large number of remote workers as small, paid tasks. These tasks, called "human intelligence tasks" or "HITs," can normally be completed in a small amount of time (seconds to minutes), and workers earn a few cents to a few dollars for each, depending upon the type and duration of the work (Ross et al., 2010).

Mechanical Turk supports a type of crowdsourcing that is sometimes referred to as "microwork" or "distributed human intelligence tasking." Microwork platforms distribute tasks that require human cognitive abilities across a large number of people via Internet or mobile connections. Considered within Quinn and Bederson's (2011) framework, microwork lies at the intersection of human computation and crowdsourcing. Microwork platforms can support a range of projects that incorporate different building blocks from the Malone et al. model (2009). Though projects requiring independent decisions fit most easily within the microwork classification, collection and collaborative creation can also be supported by these systems.

Microwork projects utilize a top-down task-assignment strategy. The initiator of the project either designs a system specifically to collect the human intelligence actions desired or defines a series of HITs for completion within an existing platform, like mTurk. In both cases, there is a leader who generates the project and workers who complete the tasks. This type of crowd work is closest to crowdsourcing's roots in outsourcing.

At the CHI 2011 workshop on Crowdsourcing[3], 21 of 43 papers focused exclusively on the Mechanical

---
[3] http://crowdresearch.org/chi2011-workshop/

Turk platform or projects that used Mechanical Turk either as a functional component of the system or for gathering research data. Seven other papers talked about different systems that supported some other form of microwork. Researchers may be drifting into this narrow view (microwork) and single instantiation (mTurk) of crowdsourcing because Mechanical Turk is easy and cheap to study—Ross et al. (2010) found that workers make about $2.00 an hour—and because Mechanical Turk can serve multiple purposes in research: i.e. a research site or object (e.g. Ross et al., 2010); as a tool for coding research data (Paul et al., 2011); and as a resource for generating classification data for machine learning algorithms. However, collectively intelligent crowd work is far more varied than microwork, and the crowdsourcing term itself, as outlined by Howe (2006, 2008), was originally offered as an explanation for a much wider variety of collective behavior.

## CHARACTERIZING CROWD WORK DURING MASS DISRUPTION EVENTS

Through my research activity as a participant-observer during mass disruption events in 2010 and 2011, I have both studied and participated in a several volunteer efforts that leverage the crowd to do information processing work. In this section, I will describe some of those efforts, and situate them with the related domains of crowdsourcing, collective intelligence, and human computation. While some of these efforts fit beneath the crowdsourcing umbrella, others are less easily characterized within existing frameworks. I intend to demonstrate that the current understanding of crowdsourcing may not be broad enough to capture all forms of crowd work during mass disruption, or specific enough to highlight the unique dynamics of information processing by the crowd within that context.

### Crisis Mapping

Crisis mapping has emerged as a popular new genre for volunteer activity during crises and mass disruption events. Generally, crisis maps are maps of an impacted area that users collectively create and edit. Volunteers, both local and remote, work to geo-locate pieces of information and put them onto a shared map. Crisis maps can be important after an event, where landmarks have changed and locations of things such as shelters or roadblocks are in flux. Goodchild (2007) describes a swell of participation in collaborative map-building web resources via volunteered geographic information (VGI), relating this to the idea of using citizens as sensors. Later research refers to these efforts as "crowdsourcing" and claims an important emerging role for them during crises (Goodchild & Glennon, 2010).

Volunteers for crisis mapping projects can come from a local community in response to a specific event, or from a growing pool of individuals who identify as "crisismappers" and repeatedly participate across events. Multiple platforms and communities have arisen to support different types of crisis mapping efforts, both reflecting and establishing different organizational structures.

*Google MyMaps*

Google MyMaps are shared maps that users can update with features, marking them with graphic icons accompanied by textual explanations. MyMaps can be started by anyone in the crowd, and there is a low barrier for entry, as no specific skills are needed for participation, just knowledge of the surrounding area or evolving conditions on the ground. For these reasons, Google MyMaps are often started by and used by locals during a crisis event. During the Fourmile Canyon Fire in Boulder, CO in 2010, two MyMaps were created by locals, documenting road closures, fire lines, donation drop-off locations and damaged structures. The most popular[4] received over a million views and had many "collaborators."

Google MyMaps generated during crisis events involve citizen reporting and collaborative creation— two forms of crowdsourcing outlined by Howe (2008). The initiation of a MyMap project resembles an open source effort: someone starts a map and then uses the web to tell the world about it and solicit help. The organization is self-organizing, with participants working together to decide what their rules for contribution are. Information is volunteered from people, some of whom may have first hand knowledge of conditions from "the ground." This type of crowdsourcing leverages the collective intelligence of a group of people to produce a resource that could be used by others who are affected or by formal responders to improve situational awareness.

*OpenStreetMap*

In the wake of the Haiti earthquake in January 2010, an existing community of mapping volunteers, OpenStreetMap, utilized their open source technology to create a collaborative map for Haiti that became a valuable source of information for relief efforts (Harvard Humanitarian Initiative, 2011).

OpenStreetMap is an international effort to create and maintain free and publicly editable maps. The mapping interface builds on the wiki concept, allowing collaborative creation. Unlike other mapping efforts discussed here, OpenStreetMap efforts are more specifically focused on geographic information than mapping changing humanitarian conditions or citizen reports. OpenStreetMap requires a more advanced skill set for participation than Google MyMaps and volunteers must be trained in specific GIS techniques in order to participate. Like Google MyMaps, their efforts can be classified as citizen sensor projects and collaborative creation. Due to the specific training required for volunteering information and the ongoing nature of mapping projects, the OpenStreetMap organization is more formal and less lateral than Google MyMaps, though still largely self-organizing. The name of the group reflects origins in the open source ideal, though the crowdsourcing term is now often applied to their efforts, and the project incorporates mechanisms of both collective intelligence, by pooling knowledge resources from a large group to generate a common picture, and distributed human computation, by deploying a large number of (trained) humans to collect multiple, discrete observations.

---

[4] Screenshot of that map: http://andrewhy.de/four-mile-canyon-fire-2010-boulderfire/

*Ushahidi*

Another example of crisis mapping is the Ushahidi platform. Ushahidi emerged from efforts by bloggers to support citizen journalists during a period of violence after a contested election in Kenya in December, 2007 (Okolloh, 2009). The platform was initially created to allow workers to assemble citizen reports of ethnic violence, arriving via SMS and the web, and to filter, verify, and map those reports. In its first deployment, Ushahidi represented an extremely rapid self-organizing effort by volunteers to connect via social media, create a tool to support their work, and develop complex work processes to maintain their ad hoc group (Okolloh, 2009).

In the early aftermath of the Haiti earthquake, an instance of the Ushahidi platform[5] was deployed, in conjunction with an SMS shortcode effort, to collect reports from affected people. Volunteers were recruited to identify actionable information, translate it into English and French, geolocate it, and structure it into a report stored within the system. The Ushahidi mapping effort continued for months, eventually transferring the responsibility for processing the reports from Tufts University students and remote volunteers to Haitian workers coordinated through Crowdflower and Samasource (Meier, 2010).

After the Haiti earthquake, Ushahidi instances were deployed for dozens of crisis events in 2010 and 2011. There was also a "formal" deploy connected to the Standby Task Force and UN-OCHA that documented violence and the movement of people during political unrest in Libya in the Spring of 2011[6], with the goal of supporting humanitarian operations (Standby Task Force, 2011). The Standby Task Force is a group of crisismappers who work together to create Ushahidi maps during events.

Supported by networks of both technical and non-technical volunteers, Ushahidi is a crowdsourced platform both in its development and its application. Its code base is open source, and is maintained by a global team of (partly volunteer) computer programmers. Within the platform, volunteers with other skills participate in a complex work process that coordinates tasks of media monitoring, report creation, report verification, and geo-locating. Meier proposed disaggregating the media monitoring and geo-location tasks of Ushahidi into human-intelligence tasks (HITs) that could be handled by workers in systems like Mechanical Turk, and indeed that model was used to sustain the Haiti Ushahidi map over time (Meier, 2010). However, my research of volunteers in action suggests that Ushahidi instances cannot be fully reduced to microwork systems, and require substantial effort by self-organizing volunteers to deploy instances, recruit workers, and coordinate efforts across a range of tasks. The crowdsourcing of Ushahidi incorporates elements of open source, outsourcing, human computation, and collective intelligence, weaving ground-up coordination efforts with the top-down task distribution of microwork.

**Technology-focused volunteer networks**

There are a few networks of volunteer programmers who focus their efforts on the crisis domain, but not strictly on mapping. Through my ethnographic investigation of digital volunteerism, I have interacted with several of these organizations during events in 2010 and 2011.

*CrisisCommons*

CrisisCommons[7] is a network of technical and non-technical volunteers focused on instigating and coordinating technical or tech-centered solutions to problems within the crisis domain. Formed in 2009, the group was originally focused on hosting events (*CrisisCamps*) that brought together volunteer programmers with domain experts for weekend "barcamps" to brainstorm, design, and develop both hardware and software solutions. In the wake of the Haiti earthquake, CrisisCommons quickly organized numerous camps to work on problems specific to that event. The organization has now established an ongoing structure between camps, and has helped to coordinate remote response with technical volunteers during subsequent events, using public wikis, Skype chats, and conference calls to organize their efforts[8].

---

[5] http://haiti.ushahidi.com/

[6] http://libyacrisismap.net/
[7] http://crisiscommons.org/learn-more/our-story/

The diverse work of CrisisCommons volunteers incorporates several different crowdsourcing techniques. It includes open source technology development, collaborative creation of information resources, and collective decision-making. Within their barcamps, their chats and their conference calls, they also use crowd-brainstorming techniques to gather ideas for how to respond to an event and what technological projects to work on. Then they congregate, in person or online, to create these tools or begin to generate informational resources.

Using Malone et al.'s (2009) classification scheme to describe this behavior, work is done both by the crowd and within an emerging hierarchy, motivations for most participants fall within the "love" and "glory" categories, work involves both creation and decision, and is done both by aggregating individual contributions and through collaboration of many members. Indeed, there is not much of the collective intelligence model that CrisisCommons does not use in its diverse work.

**Citizen Reporting**

The Ushahidi platform supports both citizen reporting and report processing using crowdsourcing techniques. There are other tools that accept citizen reports directly, and use computer software and in-house resources to process that information. For example, Washington State provides electronic forms that citizens can download, complete, and send in via email to report landslides[9]. The National Weather Service (NWS) solicits reports of weather information through both trained storm spotters and volunteers. In both these cases, project organizers use the crowd as citizen sensors, with volunteers on the ground organized from the top-down to provide info.

In other cases, citizen reporters share information laterally, without instruction, with other members of the public using available social media.

Remote data reporting—whether by "citizens" or trained volunteers—can be classified within all three crowd-work frameworks. Citizen reporting during crisis events resembles the citizen-science efforts Howe (2008) used as examples of crowdsourcing. In formal programs like the ones above, reporting can be considered a form of microwork. At the same time, the data are both a product of collectively intelligent activity and a resource with the potential to contribute to collective intelligence, and the activities of identifying information to report and communicating that information fall within the definition of human computation.

*Structuring using Microsyntaxes*

In 2010 the NWS experimented with using Twitter as a reporting channel for weather information, encouraging a microsyntax for Twitterers to use to mark location information in their tweets[10]. Information formatted in that way could be automatically aggregated and plotted onto a map without further crowdsourced processing efforts. This is similar to the Tweak the Tweet syntax, proposed for citizen reporting of crisis-related data via Twitter (Starbird and Stamberger, 2010). These Twitter microsyntaxes incorporate citizen reporting and a second layer of human computation to convert reported information into a machine-readable format. In the case of Tweak the Tweet, human computation may occur in either of two ways: when users format their first-hand information using TtT syntax, or when volunteers act to "structure" information already in the space by translating it into the TtT syntax and reposting it (Starbird & Palen, 2011).

**Self-organizing by Digital Volunteers**

Several research studies have noted that in the wake of crisis events volunteers will converge via social media, and improvise solutions to aid in response (Qu et al., 2009; Palen et al., 2009).

*Voluntweeters*

In (Starbird & Palen, 2011), we describe a group of volunteers who, after the Haiti earthquake, co-opted Twitter to use as a crisis communication channel to help in the response, searching for actionable information, verifying it, and attempting to route it to responders. Some volunteers also participated in structuring activities, moving information found elsewhere in mainstream or social media into Ushahidi reports or tweets that used the Tweak the Tweet microsyntax. As days passed, they began to connect and coordinate their efforts with other Twitterers who were tweeting for Haiti, eventually forming an interactive network of Twitter volunteers, or "voluntweeters" as some called themselves.

---

[8] CrisisCommons wikis for subsequent events:
http://wiki.crisiscommons.org/wiki/Honshu_Quake;
http://wiki.crisiscommons.org/wiki/CrisisCampNZ;
http://wiki.crisiscommons.org/wiki/Hurricane_Irene
[9] http://www.dnr.wa.gov/ResearchScience/HowTo/GeologyEarthSciences/Pages/report_a_landslide.aspx

[10] http://www.weather.gov/stormreports/

This example of ground-up and lateral self-organizing to process information during a disaster is resistant to the common use of crowdsourcing, particularly due to the absence of "sourcing." In this case, there is no person or entity—no leader—that has recognized a problem and elected to "source" it out to a distributed work force. The workforce has self-motivated, self-deployed, and over time self-organizes into an emergent response organization (Dynes, 1970). Though the work represents a form of collective intelligence, it is hard to differentiate between this type of self-organized work and pre-organized microwork within Malone et al.'s (2009) collective intelligence classification scheme.

*Humanity Road*

Interviews with voluntweeters revealed that many continued to tweet for new crisis events after the Haiti earthquake, often incorporating "crisis tweeter" or "voluntweeters" into their Twitter identity through the text in their profile. Several joined up with Humanity Road[11], a digital volunteer organization (with which I am a volunteer and participant-observer) that emerged from previous efforts to provide informational aid during disasters. Humanity Road works by appropriating available tools, including many forms of social media, to identify and distribute relevant information before, during, and after crisis events. Their work includes monitoring and filtering media and social media reports, verifying information, and integrating it into existing resources or creating new resources for affected populations and responders. The organization is almost exclusively digital, e.g. using Twitter to recruit volunteers and distribute information, Skype and shared Google documents to coordinate volunteer efforts, and a website to display resources. During crisis events, the group accepts spontaneous volunteers, quickly trains them and incorporates them into their activities and tasks. During early 2011, Humanity Road worked in conjunction with the Standby Task Force to help create "crowdsourced" maps for several disaster events.

The activities of this virtual organization are also hard to classify under the crowdsourcing umbrella, and again require several different combinations of the collective intelligence building blocks to fully describe their activities. Work is not distributed out, as in outsource. Instead, volunteers are brought into the group, where members coordinate activities and delegate tasks. There is some overlap with open source movements, as volunteers work on projects of collaborative creation, integrating information into common resources. Some work is collaborative but other work begins to look like microwork as individual volunteers are asked to repeatedly complete the same kinds of tasks. The dynamics of the organization may be better understood within existing organizational models of disaster volunteerism (like the American Red Cross) than technology-focused notions of crowdsourcing of collective intelligence.

**TOWARDS A NEW FRAMEWORK FOR CHARACTERIZING CROWD WORK**

I have presented three approaches for classifying digital crowd work, from perspectives on "crowdsourcing" (Howe 2006; 2008), collective intelligence (Malone et al., 2009) and human computation (Quinn & Bederson, 2011). Research reveals the crowd working in a variety of different ways as it responds to crises. In the previous section, I offered several examples of that crowd work and related those examples to the three perspectives above. I intended to demonstrate two things. First, the term crowdsourcing is too vague and too narrow to fully describe the various forms of crowd work that are occurring through social media during these events. And second, though most of this behavior can be classified within the existing understanding of collective intelligence, the salient features of this work, especially the self-organizing behavior, are not promoted when examining it from this perspective.

The problem with crowdsourcing is the implied meaning in the second half of that term: "sourcing." Though some of the formalized digital volunteer efforts enact some delegation of tasks and assign work "out," the self-organized efforts I describe within the voluntweeters group are not "sourced" in either the "outsource" or "open source" connotation. Initially, there was no leader recruiting volunteers to an effort. Instead, many volunteers spontaneously deployed themselves via digital technology and began to act in whatever ways they could think of to help. Tasks eventually developed, but they rose from the bottom up. Coordination was ad hoc and lateral.

The examples of the diverse work of CrisisCommons illustrate how the framework for collective intelligence offered by Malone et al. (2009) does not help us capture the salient features of crowd work during crises. The crowd work of CrisisCommons falls all over that collective intelligence classification scheme. Work is completed by both the crowd and by the in-group hierarchy. They do it for love, glory, and for a few, money. They participate in independent collection and collaborative creation. They vote in some cases and work towards consensus in others. At the same time, while being extremely diverse from a collective intelligence perspective, the group is really

---
[11] http://www.humanityroad.org/

only participating in one kind of activity: organizing information.

This statement is true for all of the examples of crowd work that I present here. These "systems" do integrate several techniques of crowdsourcing or online collective intelligence. The large majority blend crowdsourcing and human computation to complete very specific kinds of informational tasks, though in some cases they are self-organized groups that resist classification within either of those frameworks. In every case, within the context of disaster, crowd work takes the form of *organizing information* and *organizing people* to organize information. This research points to an alternative framework for crowd work that focuses on these dimensions, examining how people are organized (bottom-up, top-down, laterally) and how information is organized (how information is transformed to states of increased organization through the micro-actions of the crowd).

Clearly, there is room for more exploration on how to describe and define the complex, crowd-powered and crowd-leveraging processes, facilitated by social media, that are currently helping to organize information during disasters.